# The neural correlates of image texture in the human vision using magnetoencephalography

Elaheh Hatamimajoumerd [1], Alireza Talebpour [1]

[1] Shahid Beheshti University; eh.hatami@gmail.com

Correspondence should be addressed to Elaheh Hatamimajoumerd: eh.hatami@gmail.com

## Abstract

Undoubtedly, textural property of an image is one of the most important features in object recognition task in both human and computer vision applications. Here, we investigated the neural signatures of four well-known statistical texture features including contrast, homogeneity, energy, and correlation computed from the gray level co-occurrence matrix (GLCM) of the images viewed by the participants in the process of magnetoencephalography (MEG) data collection. To trace these features in the human visual system, we used multivariate pattern analysis (MVPA) and trained a linear support vector machine (SVM) classifier on every timepoint of MEG data representing the brain activity and compared it with the textural descriptors of images using the spearman correlation. The result of this study demonstrates that hierarchical structure in the processing of these four texture descriptors in the human brain with the order of contrast, homogeneity, energy, and correlation. Additionally, we found that energy, which carries broad texture property of the images, shows more sustained statistically meaningful correlation with the brain activity in the course of time,

Keywords: GLCM feature; Human vision, Computer vision, Multivariate pattern analysis; MEG; Texture.

## Introduction

Our brain recognizes objects effortlessly and rapidly in a fraction of a second [1]. Numerous studies have been conducted to reveal the neural mechanism behind this process [2-5]. Researchers utilized neuroimaging techniques to investigate the neural processes underlying object recognition across time and space in the human brain [1,6-9]. Using go/no go categorization task and event-related potential analysis on electroencephalography (EEG), Thrope et al [10] revealed that object recognition and categorization occurs at 150 ms. Isik et al [11], applied multivariate pattern analysis (MVPA) on MEG data to study the dynamics of size and position-invariant object recognition in the ventral visual stream. Their results indicate that while recognizing the objects starts early at 60 ms, Size- and position-invariant information are resolved around 125 ms and 150 ms, respectively. Cichy et al [6], investigated the spatiotemporal neural information encoded in the objects by combining the brain activities collected from MEG and functional magnetic resonance imaging (fMRI) data. They employed MVPA to calculate the time course of object processing in the human brain. Based on their findings, the time courses corresponding to single image decoding peaks earlier than time course of decoding category membership of individual objects. This implies the representations of individual images related to their low-level feature processing are resolved earlier in the human brain than category-specific information required in the visual ventral pathway.

Despite the human's brain fascinating feat in object recognition, this task remained a challenge in machine vision. Many computer scientists applied or proposed different visual features to discriminate between different objects. These features contain a broad range from low-level properties of images to mid and high-level features [12-14]. Textural property of an object and its surrounding areas is a powerful tool to describe the appearances of objects [15]. Texture-specific features [16] compute a gray level co-occurrence matrix (GLCM) of each image to extract second order statistical texture features including contrast, homogeneity, energy, and correlation for object recognition and texture analysis purpose. These features have been employed in various object recognition application such as Indian sign language recognition [17], brain tumour and breast cancer detection [18-20], and agricultural object-based image analysis [21].

Apart from that, researches have been studying the human visual system as a desirable system to develop an object recognition model which acts similar to the human's visual system. The HMAX model [22,23] and deep neural networks (DNN) [24] are two examples of successful models that were developed by the idea that a hierarchical architecture of brain areas mediate object recognition along the ventral stream. Cichy et al [25] traced the temporal representation of objects in the human brain provided by MEG Signals with their representations in a deep neural network. Their results indicate that the DNN captured the hierarchical process of object processing across the ventral stream pathway in time.

Investigating the MEG data collected from a human object vision on one side, and extracting the textural features from the stimuli on the other side, we traced the neural signature of texture-specific features over time. In the present study, we applied multivariate pattern analysis on MEG data and compared the texture representation of the stimuli (in terms of contrast, homogeneity, energy and correlation) with the representation of brain activity over time. We found the time course representing the temporal signature of contrast peaks earlier than other three feature. The peak latency increases with the order of contrast, homogeneity, energy and correlation. Moreover, energy, which captured the overall texture information, have a more sustained correlation with the neural data over time. We also traced the neural representation of HMAX layers as a successful hierarchical object recognition model. HMAX is developed by Serre et al [22] and models the feedforward object recognition in the ventral stream is based on Hubel & Wiesel model [26]. This model is based on the sweep of visual information from retina to the LGN, sent to early visual cortex, V2 and V4 and IT [27-29]. To better illustrate the hierarchical processing of textural features in the human vision, we then perform hierarchical clustering analysis on the time series extracted from the correlation of HMAX and GLCM representation of the stimuli which confirms the of processing hierarchy of these features.

## Materials and Methods

### 2.1. Dataset

The data used in this study is based on the MEG data acquired from an object recognition experiment designed and conducted by Cichy et al [6]. While all details of the experiment and data collection are reported in [6], we briefly describe the experiment in this section. During the experiment, sixteen healthy right-handed participants (10 females, age: mean ± SD = 25.87 ± 5.38 years) with normal or corrected to normal vision participated in two sessions of MEG recording and completed 10 to 15 runs of experiment per session. In each run of the experiment, a sequence of 92 real-world stimuli was shown to the participants in a random order. These 92 stimuli are selected from six different categories including human and non-human body, human and non-human faces, natural and artificial objects and each stimulus displayed twice in each run of the experiment.

## 2.2. MEG signal pre-processing

MEG data were recorded from 306 sensor channels (Neuromag, Triux, Elekta, Stockholm). We utilized Maxfilter software (Elekta, Stockholm) for the possible head movement correction. Afterwards, we extracted MEG trials from 150ms pre-stimulus to 1000ms after the image onset using brainstorm software [30]. The eye blink artifacts were detected and removed through the MEG frontal sensors. In the next step, we discarded and rejected the trials having a peak-to-peak value greater than 6000 fT and considered them as bad trials. To denoise the data, we finally applied a low-pass filter with a cut-off frequency equals to 30 Hz on the remaining trials.

## 2.3. Multivariate pattern analysis (MVPA)

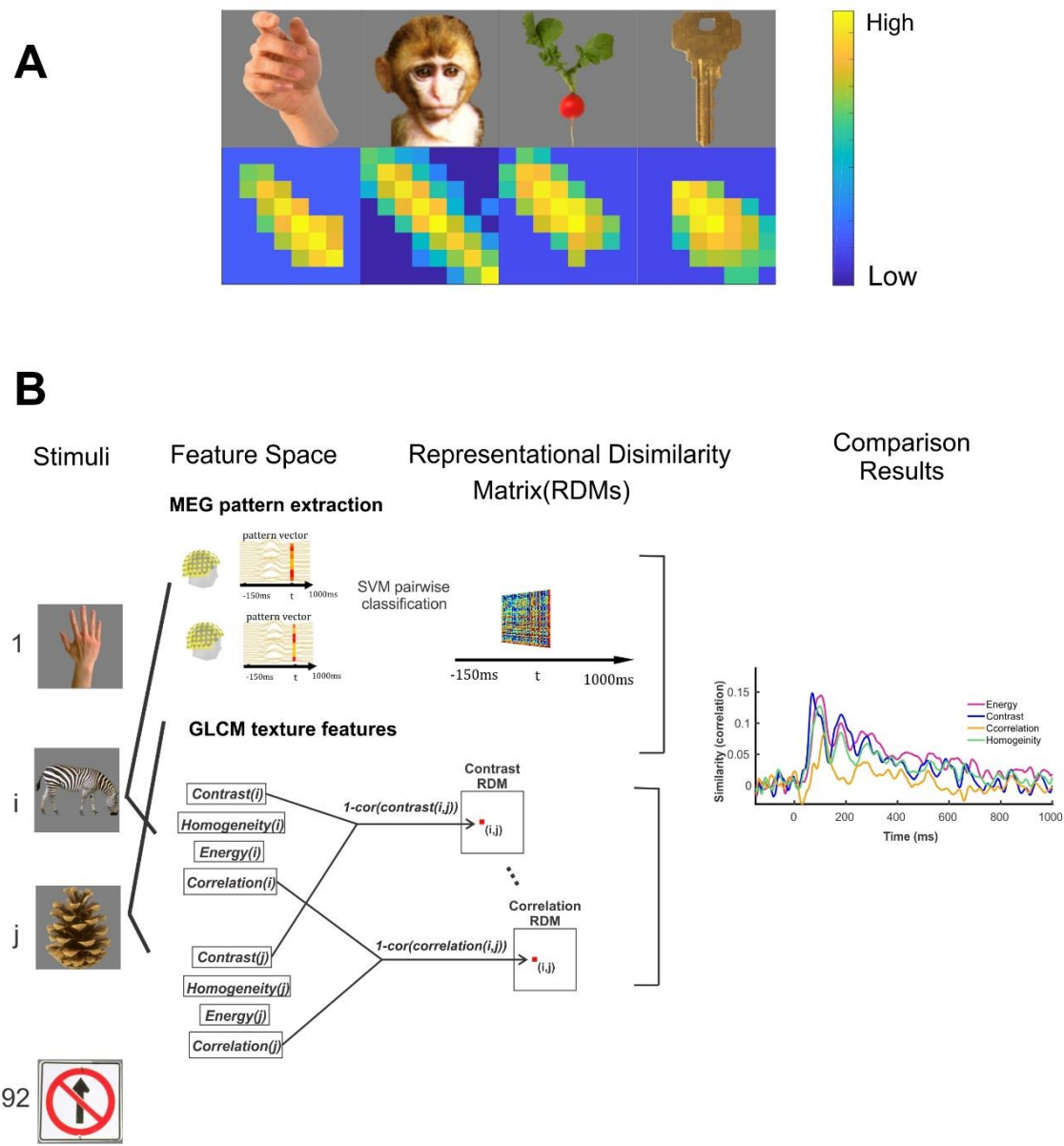

Figure 1 a) Some examples of images and their visualized GLCM matrices using eight level of gray level intensity b) Data analysis pipeline of the present study using multivariate pattern analysis and GLCM feature extraction.

We exploited the multivariate pattern classification method [31-33] to decode the objects neural information encoded in the MEG data. Figure 1b describes the pipeline analysis of this study. As depicted, we modelled the perceptual difference between each pair of stimuli in both MEG pattern space and GLCM texture features and then, stored them in representational dissimilarity matrices (RDMs). Considering 40 trials per stimulus and the 306 MEG channels, in each time point we have forty 306-dimensional MEG pattern vectors for each stimulus. In order to estimate the decoding accuracy between each pairs of stimuli, first, we permuted the order of their corresponding trials and averaged every 10 trials sequentially which resulted 4 trials per stimulus condition. Then, we trained a linear binary SVM classifier on each time point (millisecond) using 3 trials per stimulus and held the remaining one for testing the classification accuracy. To have a better assessment of decoding accuracy, we repeated this procedure (permutation, averaging and training the classifier) 100 times. The averaged of classification accuracies over the hundred repetitions is calculated as a dissimilarity measure. Having 92 stimulus, by the end of this process, a 92×92 representational dissimilarity matrix (RDM) is generated for each time point in which the element in row i and column j illustrates the decoding accuracy of SVM classifier for stimuli i and j at that time point.

## 2.4. Gray level co-occurrence matrix (GLCM)

Many computer vision applications extract the second order statistical texture information from images using gray level co-occurrence matrix (GLCM). A GLCM is a matrix with the number of rows and columns equals to the number of the image gray levels N [34]. Considering an N×N co-occurrence matrix Pij where each element (i,j) of this matrix represents the frequency of occurrence of two pixel with gray level i and gray level j located in specified spatial distance(offset).The first row of Figure 1A shows some examples of images and the second rows of Figure 1a represents their corresponding visualized GLCM matrices using eight level of gray level intensity. Afterwards, we computed four features including contrast, correlation, energy and homogeneity. Local contrast property of stimuli is defined in Equation 1. Homogeneity, measuring the closeness of distribution of the GLCM values of a stimuli to the GLCM diagonal, is represented in Equation 2. Energy, calculated by the sum of squared elements in the GLCM, is defined in Equation 3 and captures the overall textural information embedded in GLCM. Equation 4 defines the correlation as forth texture-related feature which Measures the joint probability occurrence of the specified pixel pairs.

$$Contrast = \sum_{i,j} |i-j|^2 p(i,j) \qquad (1)$$

$$Homogeneity = \sum_{i,j} \frac{p(i,j)}{1+|i-j|} \qquad (2)$$

$$Energy = \sum_{i,j} p(i,j)^2 \qquad (3)$$

$$Correlation = \sum_{i,j} \frac{(i-\mu_i)(i-\mu_j)p(i,j)}{\sigma_i \sigma_j} \qquad (4)$$

## 2.5. Statistical analysis

Non-parametric statistical tests [35,36], including signed permutation tests and cluster-wise multiple comparison correction methods were employed to estimate the significant time points of the time courses resulted from the Spearman correlations between the texture representations of stimuli and object decoding accuracies using MEG time series. According to this method of testing, permutation was used as a procedure to determine an empirical distribution. Since all the time series represent the correlation values, the null hypothesis is set to zero. The number of permutation and cluster defining threshold and cluster size correction thresholds were set to 1000, 0.05, and 0.05, respectively.

We used bootstrapping to test and estimate the peak and onset latencies of the time courses. The time series for each subject were bootstrapped and averaged across the subjects 1000 times. The standard error of measurement (SEM) is defined based on the distribution of obtained peaks of all bootstrap samples. SEM [37], is defined as $\sigma_M = \sigma/\sqrt{N}$ where σ denotes the standard deviation of the bootstrap sample distribution of the mean and with the sample size N=16 (the total number of subjects).

## 2.6. HMAX representation of stimuli

The HMAX model is a biologically inspired method used on machine vision for object recognition. In this study, we employed the HMAX model developed by Serre et al [22] (http://cbcl.mit.edu/software-datasets/index.html) imitating the hierarchical structure of visual cortex [28]. This model comprises four layers including two simple (S) and two complex (C). Each S layer is followed by a C layer. In S layers, an input image is convolved with pre-specified filters such as oriented lines. Then, the output is passed to C units which select the maximum and make the model invariant to the object size and location [38].

## 2.7. hierarchical cluster analysis

we performed agglomerative hierarchical clustering [39] on a set of eight timeseries representing the HMAX and GLCM features signatures in the human brain to analysis the similarity between the timeseries and visualize the merging clusters. This method initially considers every time series as a separate cluster. Using bottom-up approach, this method takes a distance matrix as an input and builds a hierarchy of clusters by merging the most similar (closest) clusters based on a proper linkage criterion. In this study, we used the complete-linkage[40,41] criterion defined in Equation 5 in which d calculates the distance between every item a in cluster A with each item b in cluster B. based on this measure, the distance between every two cluster A and B is the is the maximum distance between the time series in cluster A and cluster B. The hierarchical clustering process can be visualized by a dendrogram [42].

$$d(A, B) = max(d(a, b): a \in A, b \in B \quad (5)$$

## Results

### 3.1. The temporal signature of GLCM texture-related features in the brain

To trace the signature of texture-specific features in the human visual system, we first, extracted four visual features including contrast, homogeneity, energy and correlation from the gray level co-occurrence matrix of each stimulus. These visual features capture the overall textural properties of the stimuli. Apart from that, we also applied multivariate pattern analysis to create time-resolved representational dissimilarity matrices from MEG patterns corresponding to image stimuli. As depicted in Figure 1b, in both MEG and textural feature spaces, we performed the representational similarity analysis (RSA) to calculate the RDMs over the time as well as four texture-related (contrast, homogeneity, energy and correlation) RDMs (Figure 2b). Since the 92×92 RDMs in both spaces carry the pairwise distances of stimuli, we performed Spearman correlation between each of texture-related RDMs and MEG

RDMs over the time to trace the similarity of textural representational of stimuli with the neural data over the time. The time courses corresponding to each texture-related feature and brain data is shown in Figure 2a. color-coded lines above the curves shows the significant time points. All the significant time points are estimated using non-parametric permutation statistical tests with cluster defining threshold P<0.05, and corrected significance level P<0.05 (N=16) (see section 2.5). As Figure 2a illustrates, considering the significant time points, energy which represent the sum square of GLCM- has a more sustained signature in brain activity than the other three features. This can be explained because the energy captures the overall texture information encoded in the images while the three features tend to represent images in more specific aspects such as contrast, homogeneity and correlation.

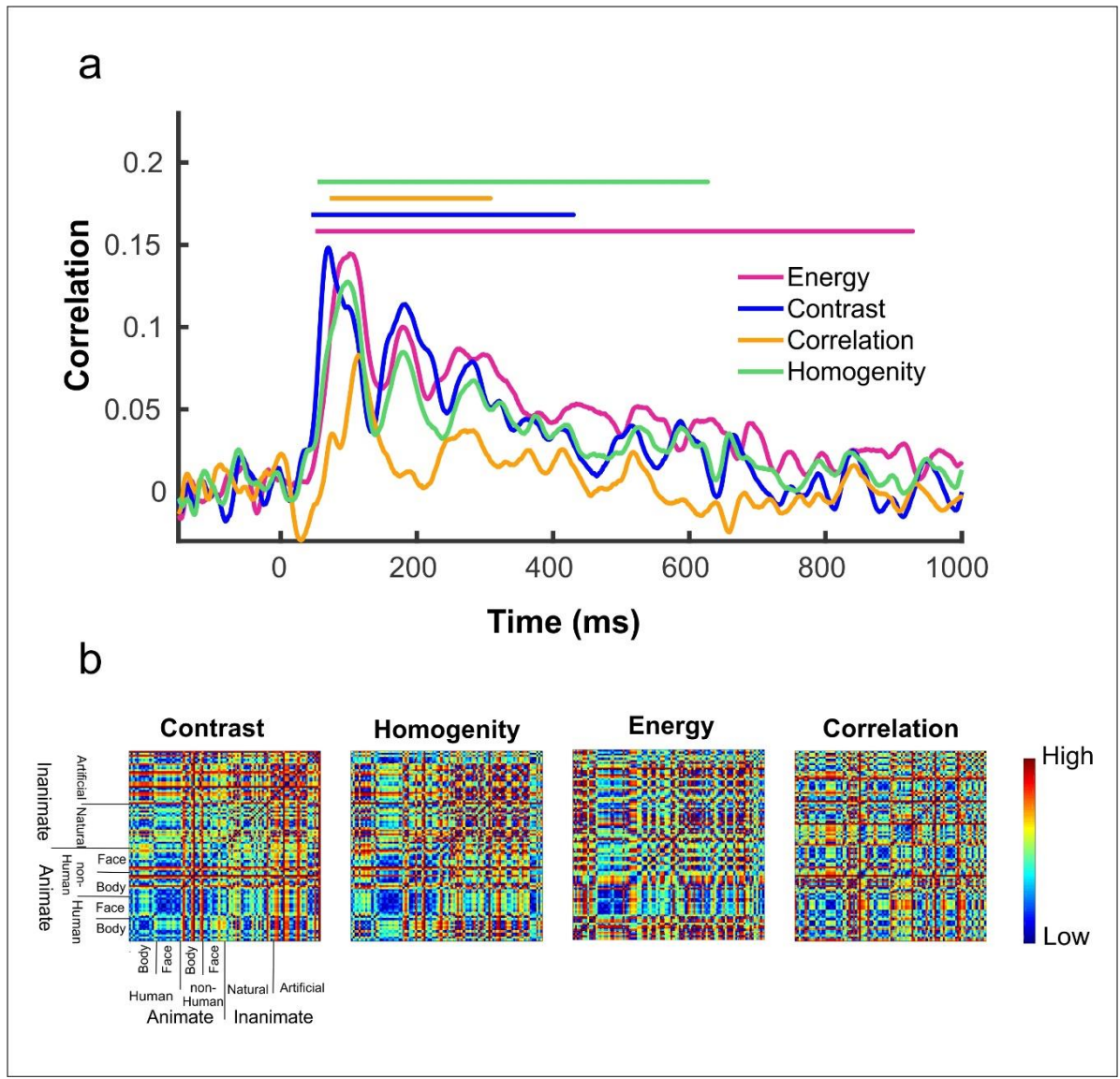

Figure 2 a) The temporal neural signature of GLCM texture feature resulted from performing the Spearman correlation of the texture feature including contrast, homogeneity, energy and correlation with the pairwise MEG decoding in each time point. b) The RDMs of GLCM feature (from left to right: contrast, homogeneity, energy and correlation) representation of stimuli.

## 3.2. Peak latencies analysis of texture-related Time courses

Knowing the information provided in previous section, we calculated the peak latencies of all the time courses to investigate if the categorical information is conveyed by these features. Figure 3a, b depicts the peak latencies of the time courses corresponding to texture-related features (contrast, homogeneity, energy, correlation). As can be seen in Figure 3a, the peak latency increases monotonically from Contrast to Correlation. The peak latencies are evaluated using 1000 bootstrap with sample size N=16 over the participant Time courses. Error bar indicates the standard error of the measurement (SEM).

The increasing trend of peak latency suggests that there is more categorical information embedded in Energy and correlation representation of stimuli as they are processed in later regions. To better visualize the distribution of peak latencies for these textural features, we presented the histograms of the peak latencies in Figure 4. The histograms are generated using 1000 bootstrap sampling of the subject-specific time courses for each feature. The red curves on this figure shows the normal distribution fitted on the histograms. The result of Figure 4 confirms the previous information extracted from Figure 3.

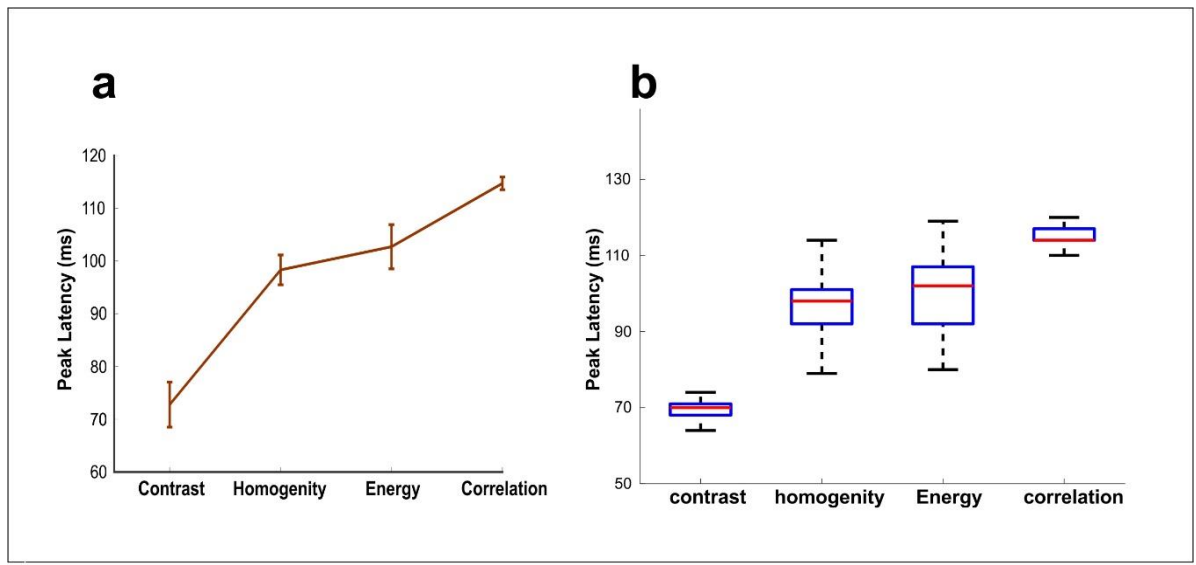

Figure 3. Peak latencies of the time courses corresponding to texture-related features contrast, homogeneity, energy, correlation) represented by (A) Error bar and (B) Box plot. Peak latencies are evaluated by signed permutation test non-parametric permutation statistical tests using cluster defining threshold P<0.05, and corrected significance level P<0.05 (N=16). Error bar indicates the standard error of the measurement (SEM) calculated by 1000 bootstrap with sample size N=16 over the participant timecourses.

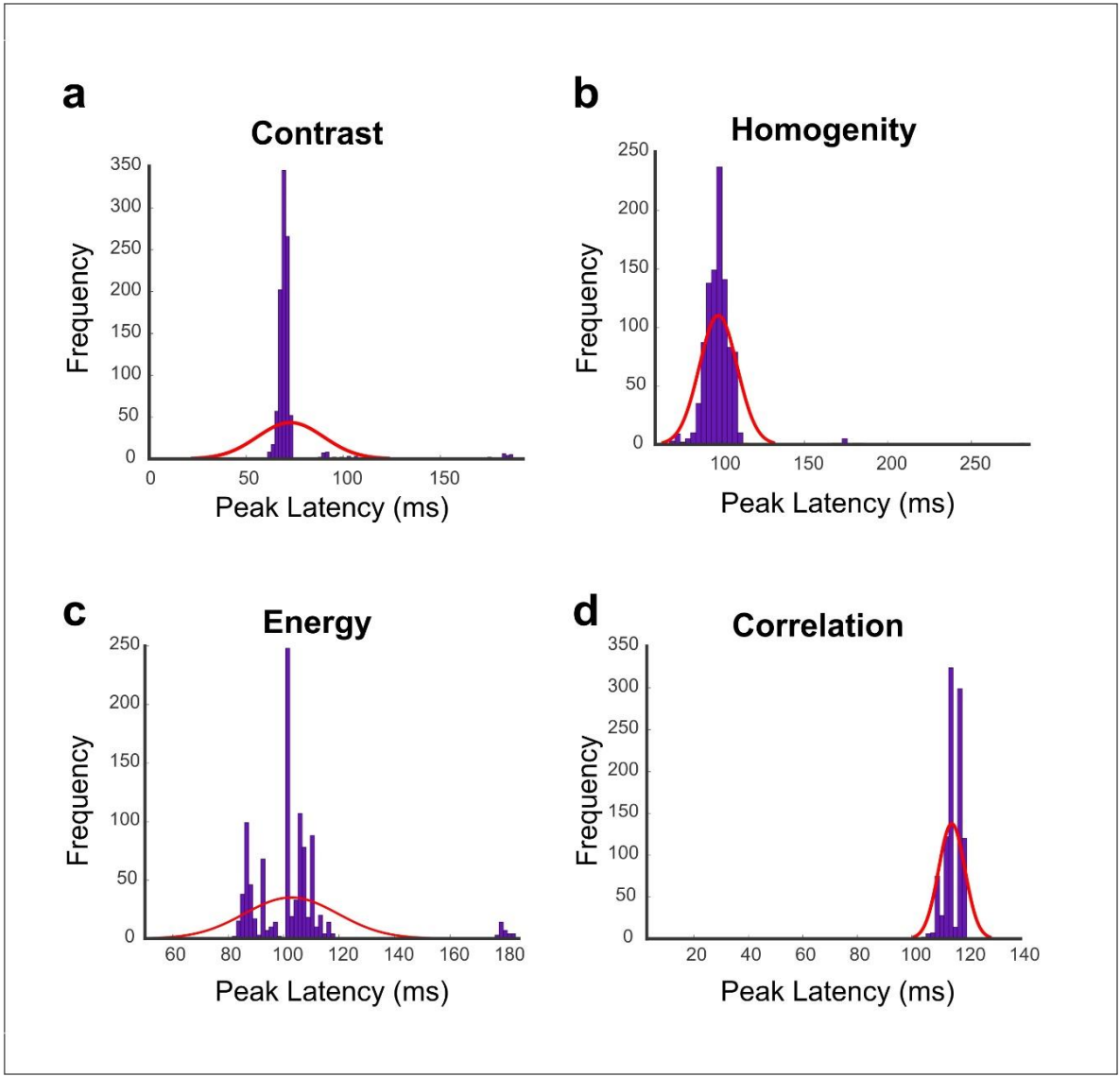

Figure 4 Figure 4 The histogram of peak latency generated on 1000 bootstrap samples with sample size N=16 for A) Contrast B) Homogeneity C) Energy and D) Correlation. The red curves illustrate the Normal distribution fitted on the histogram data.

### 3.3. Hierarchical clustering between HMAX and GLCM textural feature

To better examine the GLCM feature processing in the human brain, we used HMAX, a renowned biologically-inspired hierarchical object recognition model, as a benchmark and performed the hierarchical cluster analysis on a set of HMAX and GLCM timeseries. Figure 5a illustrates the signature of HMAX layers in the human brain calculated by the Spearman correlation between the MEG decoding time course and every HMAX RDM (Figure 5b). The similarity matrix representing pairwise correlation between GLCM and HMAX feature is depicted in Figure 6b. The dendrogram in Figure 6a illustrates the sequential bottom-up merging clusters. As depicted HMAX1 and HMAX2 containing the low-lever features remained merged from the beginning of process and the GLCM features tend to be merged with HMAX3 and HMAX4. While Contrast is closest to HMAX3, Homogeneity is merged with HMAX4. In the next step, Energy is added to HMAX4 and Homogeneity. This process shows that there is a hierarchy in process and Contrast is processed earlier in visual cortex.

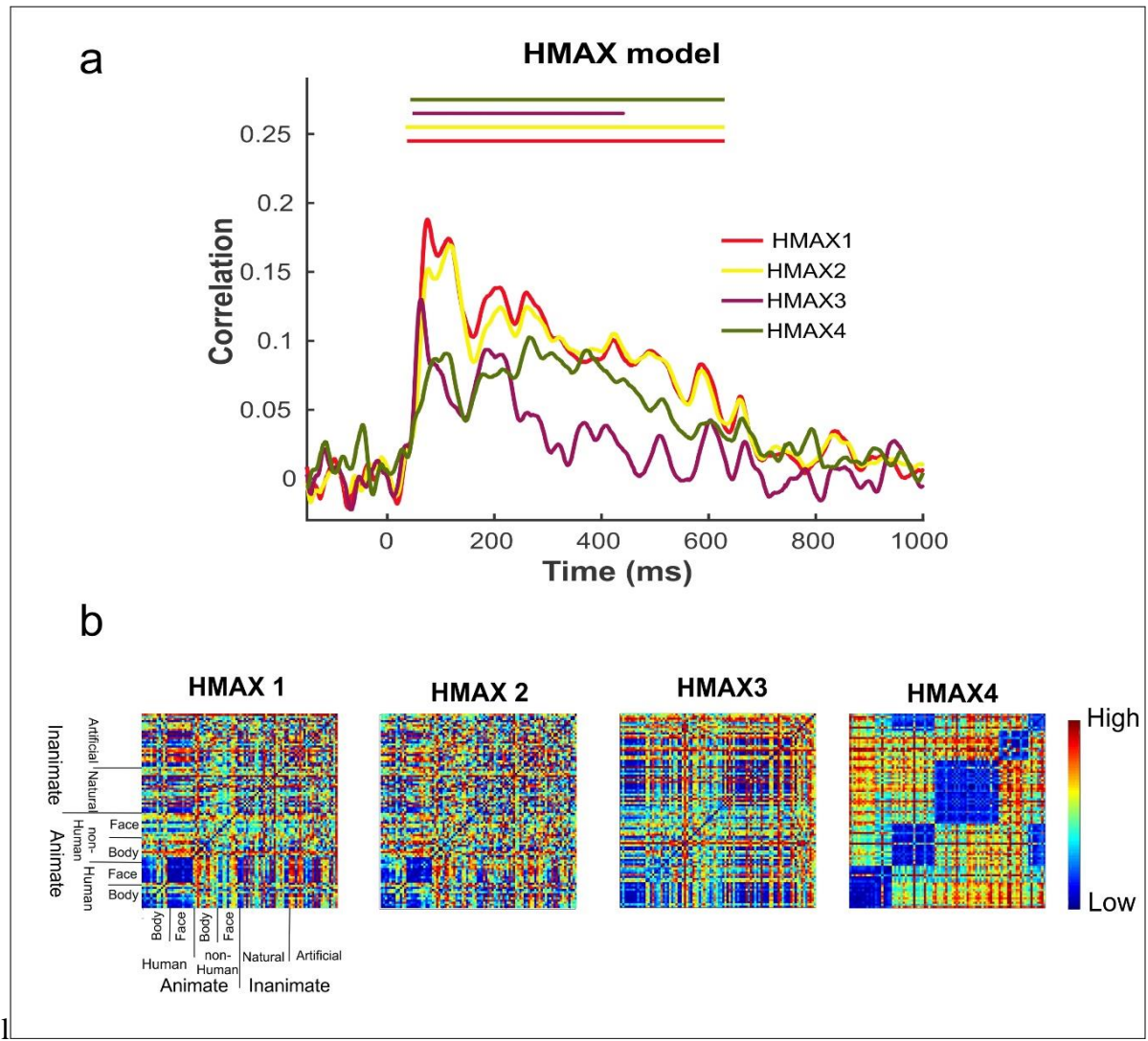

Figure 5 a) The temporal neural signature of HMAX model resulted from performing the Spearman correlation of four layers of HMAX with the pairwise MEG decoding in each time point. b) The RDMs of HMAX layers (from left to right: HMAX1, HMAX2, HMAX3 and HMAX4) representation of stimuli.

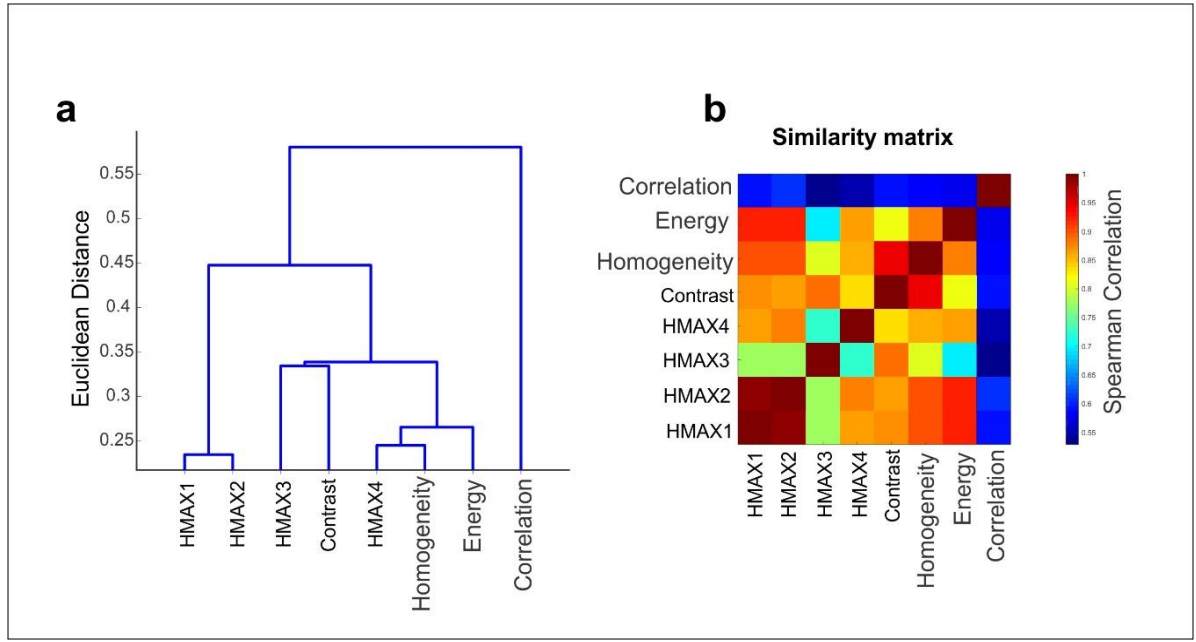

Figure 6 a) The dendogram resulted from agglomerative hierarchical clustering b) correlation matrix representing the pairwise spearman correlation between GLCM and HMAX timecourses.

## 4. Discussion

### 4.1. Summary

We applied multivariate pattern analysis on MEG data and representational similarity analysis on textural features extracted from the stimuli to trace the neuro-dynamic of textural information in the human brain. Our result illustrates that there is a sequential processing for these four textural features (contrast, homogeneity, energy and correlation).

### 4.2. The peak latencies of the time courses represent the decoded visual information of stimuli

Modelling the object recognition across the ventral stream pathway yielded several applications in image processing and computer vision. Cichy et al [25] compared the DNN layers features of the stimuli with their MEG representation. They found that the peak latencies corresponding to the correlation time courses of the first layers of deep neural networks occurs earlier than last layers which confirms the swept of information across the ventral stream from processing the low-level to mid and higher-level semantical properties of the stimulus. As Figure 2 illustrates, the peak latencies of the correlation time courses between four texture-related features and MEG RDMs occurs at 70, 99,102 and 114 ms after the stimulus onset for contrast, homogeneity, energy and correlation features which shows a hierarchy in the timing process of textural features in the human brain. This result indicates that, while contrast and homogeneity carry mostly the low-level features, energy and correlation contain more semantic and higher-level information that needed to be processed in the later area across the ventral steam. To better visualize the hierarchical processing of textural features in the human vision, we perform a comparative study of GLCM features and HMAX model using hierarchical clustering analysis. Since HMAX is a biologically-inspired model of object recognition by hierarchical and feedforward processing of low, mid and high-level information [22], it can serve as a benchmark for this study. hierarchical clustering analysis on the time series extracted from the correlation of HMAX and GLCM representation of the stimuli (Figure 6a) demonstrates that while Contrast feature tend to merge with HMAX3, Homogeneity merges with HMAX4. Then Energy is added to the cluster containing HMAX4 and Energy and finally the correlation feature is added to them. This process shows the hierarchical processing of these textural features.

In various studies texture has been considered as a low-level property of an image required to be processed in early visual cortex (EVC) in the human brain [43]. Gabor filters are known as a model the of EVC neurons [22,44], in processing of low-level features such as edges and also texture analysis. In this study, we extracted four texture-related features to capture the different aspects of textural information using the GLCM matrix. Since the GLCM contains the two-dimensional gray level variation information, it can capture the mid-level feature so, it models the textures better than Gabor filters which considered a low-level feature. Considering that the object recognition process can started as early as 60 ms after the stimulus onset, and higher-level categorical information processed later stages of ventral stream about 150 ms [10,11], the peak latencies of texture-specific features reported in Figure 3 shows that these features represent the low and partially high-level properties of the images which is critical to the object recognition and categorization process.

Apart from that, these features have been widely used in computer vision to capture the texture-specific information and successfully managed to discriminate between the classes of objects such as brain tumors [18,19] handwritten detection[17] . The performance of these feature for object recognition demonstrate their noticeable level of semantic information yielded to object recognition.  SEEMORE[45], an object recognition model in the engineering domain, consider a combination of color, shape and texture feature extraction to imitate the feed-forward object processing in the brain.  They computed the texture representation of image by the variances of the oriented energy, measured by the mean squared deviation of the total energy at different orientations. Considering the significant time point in Figure 2 depicted by the color-coded horizontal lines above the curves, our results confirm that the energy have a more sustained meaningful correlation with the neural data over the time than other three textural features ($p<0.01$). this can be explained as the energy (calculated by sum of squared elements in the GLCM) captures the overall texture information of the stimuli which contains the broad range of low mid and high-level information.

## 5. Conclusions

We investigated the neural signature of four texture-related features including contrast, homogeneity, energy, and correlation over the time. Our results indicate that there is a meaningful correlation between representation of images with GLCM texture features and the MEG data over the time.  Furthermore, studying the peak latencies of the correlation time-courses corresponding to these four textural features demonstrates hierarchy of the timing process of textural features in the brain with the order of contrast, homogeneity, energy, and correlation. Therefore, it can be inferred that the energy and correlation contain higher level visual properties of the stimuli.

**Author Contributions:**

E.H.  and A.T conceptualized the research. E.T analyzed the MEG data and stimuli and prepared the original draft of manuscript. A.T. assisted with the interpretation of results and supervised on analysis and writing process.

**Funding:** This research received no external funding supporting this research

**Acknowledgments:** We would like to express our great appreciation to Dr. Dimitrios Pantazis and Dr. Aude Oliva who provide the MEG and stimuli dataset and provided us their advice and assistance in general MEG data analysis that greatly assisted the research.

**Conflicts of Interest:** The authors declare that the research was conducted in the absence of any commercial or financial relationships that could be construed as a potential conflict of interest.